\documentclass[aps,prb,preprint,showpacs,superscriptaddress,28pt]{revtex4}
\usepackage{graphicx}
\usepackage{amsmath}
\usepackage{bm}

\begin{document}

\title{Proximity effect can induce the energy gap rather than superconducting pair potential}
\author{Z. W. Xing}
\affiliation{National Laboratory of Solid State Microstructures and
Department of Materials Science and Engineering, Nanjing University,
Nanjing 210093, China}
\author{Mei Liu}
\author{}
\affiliation{Department of Physics, Southeast University, Nanjing
210096, China}
\date{\today}
\begin{abstract}
For an $s$-wave superconductor/semiconductor/ferromagnetic-insulator
structure, the proximity effect can induce the energy gap in the
semiconductor rather than the superconducting pair potential of its
microscope Hamiltonian. As a result, it is questionable to realize
topological superconducting states in that structure.
\end{abstract}
 \maketitle

The experimental realization of Majorana fermions has triggered an
avalanche of research activity, because such excitations satisfy
non-Abelian statistics which form a centerpiece in recent proposals
for topological quantum computations. In a recent Letter, Sau $et$
$al.$~\cite{Sau} proposed that a topological superconducting phase
supporting Majorana fermions can be realized using  a semiconductor
with Rashba spin-orbit coupling, sandwiched between an $s$-wave
superconductor (S) and a ferromagnetic insulator. The basic idea is
that the ferromagnetic insulator produces a Zeeman field $V_z$
perpendicular to the semiconductor, which separates the two
spin-orbit-split bands by a finite gap.  For $\mid\mu\mid < \mid
V_z\mid$, i.e., the Fermi level $\mu$ lies inside this gap,  only a
single band crosses the Fermi level, which is analogous to the
surface state of a strong topological insulator (TI).~\cite{Kane}
There will appear non-abelian topological order if $s$-wave
superconductivity is induced in the semiconductor by the proximity
effect. In this system, the superconducting pair potential $\Delta$
is a key ingredient of driving the semiconductor into a topological
superconducting state. However, we argue here that $\Delta$ assumed
in the semiconductor [Eq.\ (2) of Ref.\ 1] is questionable, even
though the proximity effect can induce the superconducting order
parameter and the energy gap (not the superconducting energy gap)
there.

It is of great importance to distinguish pair potential $\Delta$
appeared in the microscopic Hamiltonian from the energy gap and
superconducting order parameter. The pair potential stands for the
number of Cooper pairs in the condensate, and it determines the
basic features of superconductivity such as the Meissner effect and
vanishing electric resistance. For an S/normal-metal (N) bilayer,
both superconducting order parameter $F(z)$ and pair potential
$\Delta(z)$ always exist on the S side, which satisfy a simple
relation of $\Delta(z)=\lambda^* F(z)$ with $\lambda^* $ the
effective pair interaction.~\cite{3} On the N side, $F(z)$ can be
induced by the proximity effect, but $\Delta (z)$ determined from
the self-consistency equation equals zero because $\lambda^* $ is
supposed to be vanishing. The superconducting order parameter in the
N indicates the quantum coherence of a pair of electrons without
condensate. As regards the energy gap, it is not the mark of
superconductivity. In an S, the superconducting energy gap is the
minimal binding energy of the Cooper pair in the condensate, and the
gapless superconductivity may appear due to magnetic impurities or
the inverse proximity effect. In an N contact to an S, the
proximity-effect-induced energy gap is not a superconducting energy
gap, but a pseudo energy gap arising from the coherence of electron
pairs without condensate. Another example of the pseudo energy gap
is the normal state of the high-$T_c$ superconductors, where there
is a finite energy gap but no superconductivity. As a result,
without the BCS pair potential, the proximity-effect-induced energy
gap ought not to be viewed as a basis of assuming a finite pair
potential in the microscope Hamiltonian.

The pair potential cannot be assumed artificially, and it must be
self-consistently determined as long as it appears in the microscope
Hamiltonian and BdG equation. If both $\Delta$ and $V_z$ are assumed
to appear simultaneously in the microscope Hamiltonian, they must be
related to each other. For such an assumption, it is necessary to
show that the coexisting state for $\Delta$ and $V_z$ has the lowest
thermodynamic potential. In the absence of the spin-orbit coupling,
from the lowest thermodynamic potential principle, it has been shown
that the $s$-wave superconducting ground state can be realized only
for $0.707 \Delta > V_z$, which is the Clogston criterion at zero
temperature.~\cite{4} However, the existence of non-Abelian Majorana
fermions requires the condition of $(\mu^2 +\Delta^2)< V_z^2$, i.e.,
at least $\Delta < V_z$. Although the presence of the spin-orbital
interaction complicates the question, the lowest thermodynamic
potential principle cannot be violated.

In summary, we argue that the pair potential in the microscope
Hamiltonian of the semiconductor cannot be induced by the proximity
effect. As a result, it is questionable to realize the topological
superconducting state in the $s$-wave
S/semiconductor/ferromagnetic-insulator structure.~\cite{Sau}

This work is supported by the State Key Program for Basic Researches
of China under Grant No. 2010CB923404.

\end{document}